\documentclass[12pt,epsfig]{article}

\usepackage{cite}
\usepackage{epsfig}
\usepackage{graphics}
\usepackage{inputenc}
\inputencoding{latin1}

\def\herwig{{\sc Herwig} \hspace{.05cm}}
\def\pomwig{{\sc Pomwig} \hspace{.05cm}}

\def\C2q{C_2(Q)}
\def\tC2q{$\C2q$}
\def\f2q{f_2(Q)}
\def\tf2q{$\f2q$}

\newcommand{\PO}{I\!\!P}
\newcommand{\RO}{I\!\!R}
\newcommand{\xpom}{x_{\PO}}

\begin{document}

\sloppy


  \begin{flushright}
    MAN/HEP/2000/3 \\
    MC-TH-00/09 \\
    June 2001 
  \end{flushright}
  \begin{center}
    
    \vskip 10mm {\Large\bf\boldmath \pomwig : \herwig for Diffractive Interactions} \vskip 15mm

    {\large Brian Cox and Jeff Forshaw}\\
    Dept.~of Physics and Astronomy, University of Manchester\\
    Manchester M13 9PL, England\\
    coxb@mail.desy.de
    \vskip 10mm

  \end{center}
  \vskip 0mm
\begin{abstract}
\noindent
We have modified the \herwig event generator to incorporate diffractive interactions. All standard \herwig hard subprocesses are available. 

\end{abstract}


\section{Introduction}
\label{sec:intro}

The modifications to \herwig \cite{herwig} that are necessary to incorporate diffractive collisions are relatively simple once it is noticed that pomeron exchange events in hadron - hadron collisions look very much like resolved photoproduction events in lepton - hadron collisions. In resolved photoproduction in electron - proton collisions, for example, the process is modeled by the incoming electron radiating a quasi-real photon according to a flux formula. The photon is then treated as a hadronic object with a structure function, which undergoes a collision with the beam proton. Similarly, single diffractive interactions in proton - proton collisions may be modeled by assuming that one of the beam protons emits a pomeron, again according to some flux formula, which subsequently undergoes an interaction with the other beam proton (see figure 1). \herwig will automatically choose to radiate a photon from a beam lepton if a hard subprocess is selected which requires a hadronic structure for the beam lepton. Examples would be choosing \herwig subprocess 1500 (QCD $2 \rightarrow 2$ scattering) in an electron - proton collision, or subprocess 9000 (deep inelastic scattering) in an $e^+ e^-$ collision.

All that is necessary therefore, to simulate a single diffractive interaction in a pp collision is to replace the photon flux with a suitable pomeron flux factor, and the photon structure function with a pomeron structure function, and run \herwig in ep mode rather than pp mode. The electron should then be identified with the proton which remains intact after the diffractive scattering. This process may be generalised to include sub-leading exchanges, and to perform double pomeron collisions and diffractive DIS as will be described below. 

Our philosophy for \pomwig has been to make as few changes as possible to the \herwig code. In particular, the \herwig common blocks are left unchanged, and only two \herwig subroutines have been modified. This approach has some disadvantages, in that we have left the event record unchanged. An intact proton will still appear as a lepton, therefore, and a pomeron will appear as a photon. We felt that such cosmetic inconveniences are outweighed by the overall simplicity and ease of installation and maintenance which are features of \pomwig.  

\begin{figure}
\centering
\begin{picture}(450,85)

\put( 0,0){\epsfig{file=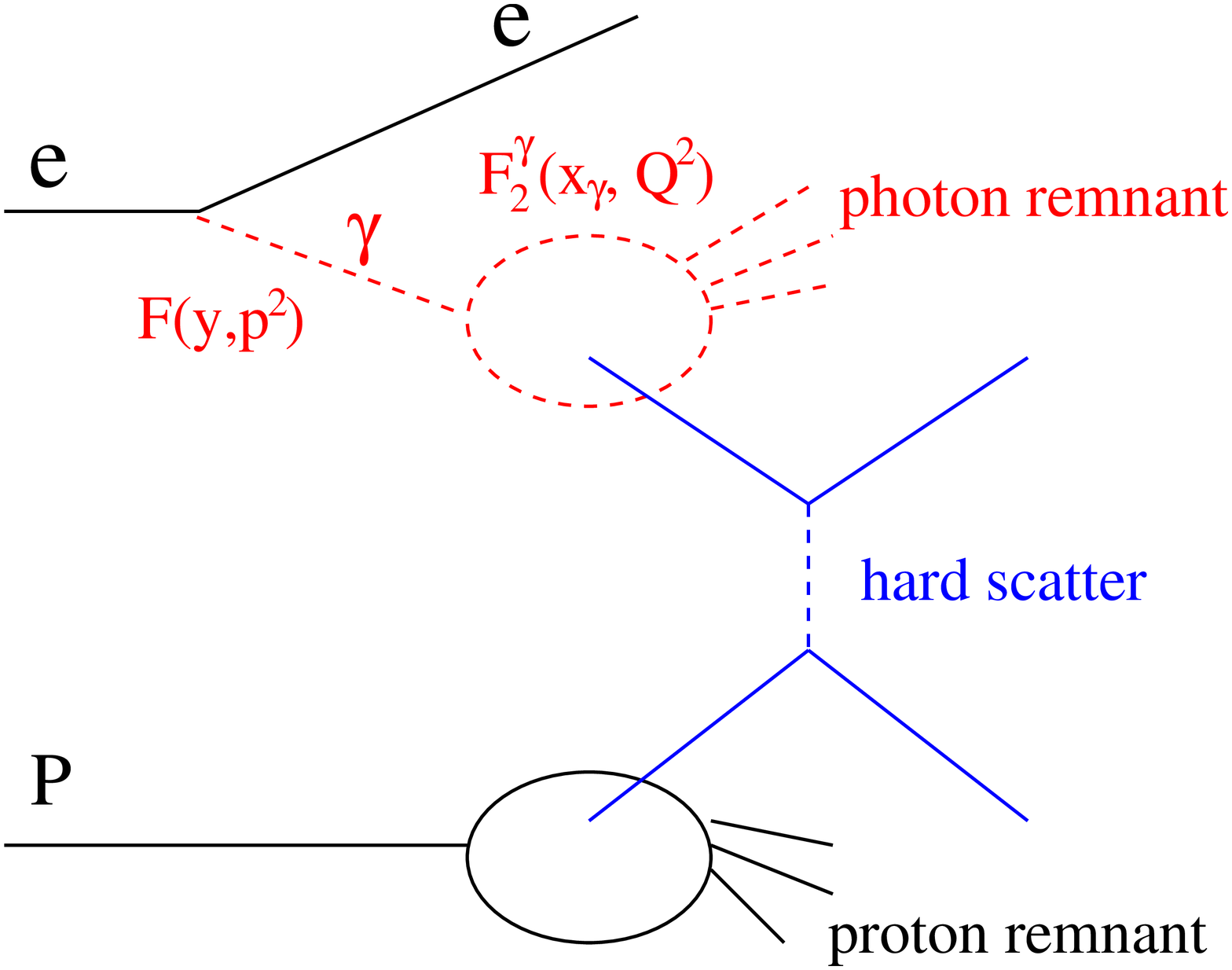,width=.45\textwidth,}}
\put( 225,0){\epsfig{file=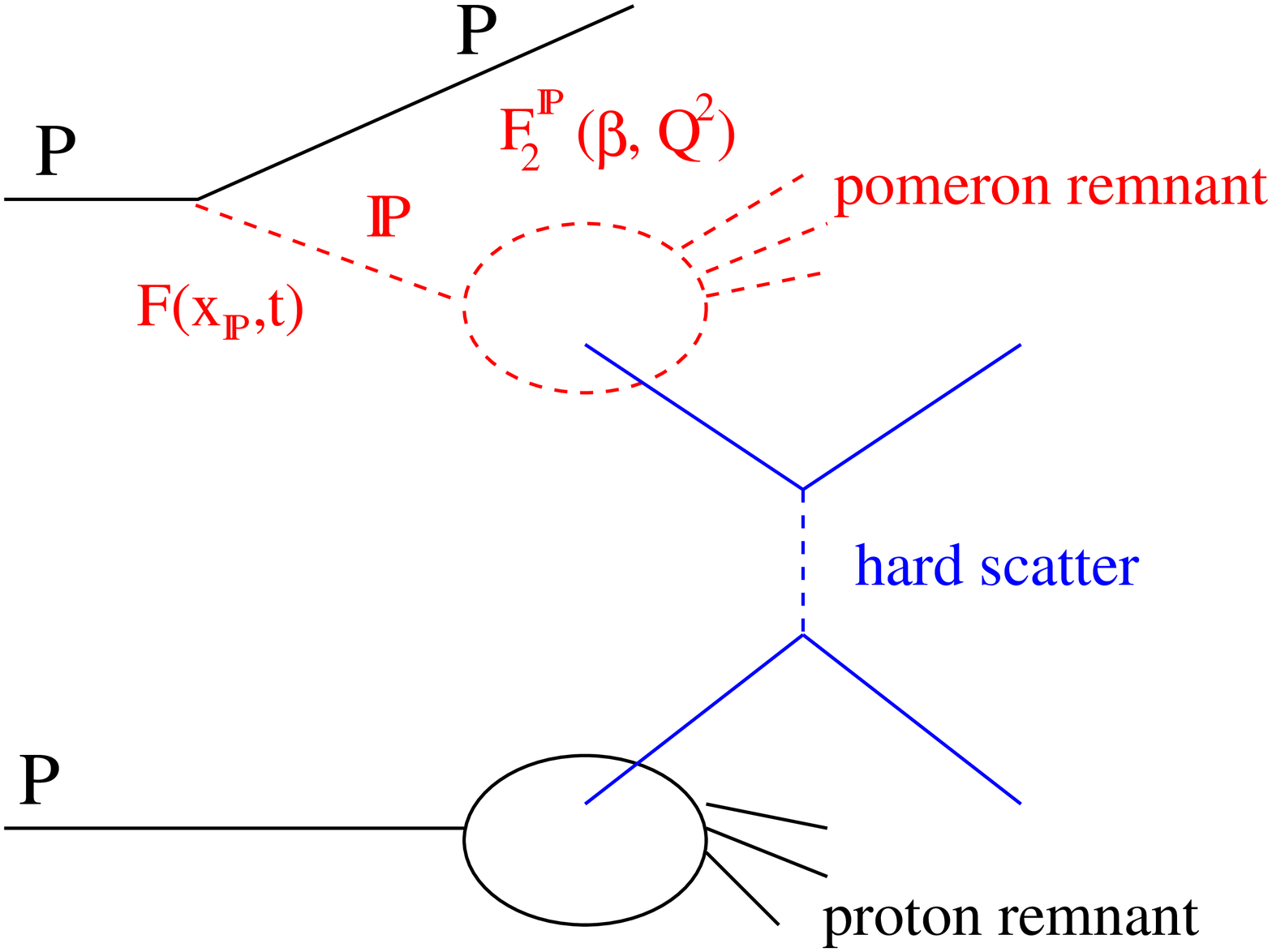,width=.47\textwidth,}}
\end{picture}
\caption{The \pomwig philosophy. Photoproduction in ep (or ee) collisions is replaced with pomeron or reggeon exchange in pp (or ep) collisions.}
\label{photopom}
\end{figure} 

\section{Installing the Code}
\label{sec:code}

The code can be obtained on request from the authors, or from \cite{pomwigcode}. The routines supplied will function with all currently available versions of \herwig from 5.9 onwards. 
There are two \herwig routines which must be replaced: \texttt{HWSFUN} and \texttt{HWEGAM}.   \texttt{HWSFUN} has been modified to call the H1 diffractive structure function routine \texttt{H1QCD} \cite{H1diff}, or the user defined structure function routine \texttt{POMSTR} in the case of pomeron exchange, or to use a neutral pion structure function in the case of reggeon exchange. The pion structure is taken from the CERN \texttt{pdflib} package via a call from \texttt{HWSFUN}. The default reggeon structure is that of Owens \cite{owens}, as used in the H1 diffractive DIS analysis \cite{H1diff}.  \texttt{HWEGAM} substitutes the H1 pomeron or reggeon flux for the photon flux. The flux itself is calculated by the new routine \texttt{FLUX}.  

If beam particle 1 is the electron (playing the role of the diffracted proton), then setting \herwig parameter \texttt{MODPDF(1) = -1} and \texttt{NSTRU = 6} will cause the H1 pomeron structure function to be used for beam particle 1. Similarly, setting \texttt{NSTRU = 7} will select reggeon exchange, and \texttt{NSTRU = 8} will select the user defined pomeron structure from \texttt{POMSTR}. The parameters in the H1 pomeron structure function routine are set via an initialisation call to \texttt{H1QCD}.

To initialise the H1 pomeron structure function, the call should be made from the \herwig routine \texttt{HWABEG} as follows; \\

\texttt{INTEGER ifit \\ DOUBLE PRECISION XPQ(-6:6),X,Q2 \\ Q2=75 \\ ifit=5 \\ X=0.1 \\ call QCD\_1994(X,Q2,XPQ,ifit)} \\

Details of the parameters can be found in the \texttt{H1QCD} routine. The above parameters select the LO fit 2. For this initialisation call to \texttt{H1QCD}, the values of \texttt{Q2} and \texttt{X} are irrelevant.
 
The \herwig parameters \texttt{Q2WWMN} and \texttt{Q2WWMX} set the minimum and maximum pomeron virtualities, $t_{min}$ and $t_{max}$ respectively. \texttt{YWWMIN} and  \texttt{YWWMAX} set the minimum and maximum incoming proton energy fractions carried by the pomeron, $\xpom^{min}$ and $\xpom^{max}$ respectively.

The default parameters for the pomeron and reggeon fluxes are those found by the H1 Collaboration in \cite{H1diff}, for the case in which no interference is assumed between the pomeron and reggeon contributions to $F_2^{D(3)}$, as shown in table 1\footnote{Note that the reggeon parameters are not well constrained by current measurements. The defaults are those used to generate the curves in figure 2.}. The fluxes are parameterised as
\begin{equation}
f_{\PO /p}(\xpom)=N\int^{t_{min}}_{t_{max}}\frac{e^{\beta_{\PO}(t)}}{\xpom^{2\alpha_{\PO}(t)-1}}
\end{equation}
\begin{equation}
f_{\RO /p}(\xpom)=C_{\RO}\int^{t_{min}}_{t_{max}}\frac{e^{\beta_{\RO}(t)}}{\xpom^{2\alpha_{\RO}(t)-1}}
\end{equation}
where $\alpha_{\PO}(t) = \alpha_{\PO}(0)+\alpha^{'}_{\PO}t$ and $\alpha_{\RO}(t) = \alpha_{\RO}(0)+\alpha^{'}_{\RO}t$. In the case of the user defined structure functions, $N = 1$ by default, and the normalisation may be included within \texttt{POMSTR} or \texttt{FLUX} at the users discretion. The normalisation of the flux is arbitrary in the case of the H1 pomeron structure function. The  \texttt{H1QCD} routine is implemented such that the generated cross section will always match $F_2^{D(3)}$ as measured by H1 at $\xpom = 0.003$, irrespective of the parameters chosen for the flux. 
\begin{table}
\begin{center}
\begin{tabular}{|l|l|l|l|l|l|} \hline

  Quantity  & Value  \\ \hline
  $\alpha_{\PO}$ & 1.20 \\ \hline
  $\alpha_{\RO}$ & 0.57 \\ \hline   
  $\alpha^{'}_{\PO}$ & 0.26  \\ \hline
  $\alpha^{'}_{\RO}$ & 0.90  \\ \hline
  $B_{\PO}$ & 4.6 \\ \hline
  $B_{\RO}$ & 2.0 \\ \hline
  $C_{\RO}$ & 48 \\ \hline
  
\end{tabular}
\caption{The default parameters in \pomwig}
\end{center}
\end{table}

\section{Summary of available processes}
The processes currently available in \pomwig are summarised in table 2. The particular hard subprocess required is set as usual via \herwig parameter \texttt{iproc}. Note that at present for double pomeron exchange, it is only possible to simulate pomeron - pomeron or reggeon - reggeon collisions, and not pomeron - reggeon. In diffractive DIS (\texttt{iproc=19000}), the pomeron is always emitted from the second beam particle. Note also that process $eP \rightarrow ePX$ in table 2 refers only to diffractive DIS. \pomwig is unable at present to generate this process with a resolved photon component.

\section{Example processes}
\begin{table}
\label{sum}
\begin{center}
\begin{tabular}{|l|l|l|l|l|l|} \hline

  Process  &  BEAM1 &   BEAM2 & MODPDF(1) & MODPDF(2) & NSTRU \\ \hline  
  $P P\rightarrow P X$ & P & $E+$ & \texttt{pdflib}  & -1 & 6 (7, 8) \\ \hline
  $P \bar P\rightarrow P \bar P X$ & E+ & $E+$ & -1 & -1 & 6 (7, 8) \\ \hline
  $ e P \rightarrow e P X$ & E+ & E+ &  N/A & -1 & 6 (7, 8) \\ \hline
\end{tabular}
\caption{The diffractive processes available in \pomwig}
\end{center}
\end{table}
In order to make the modifications clear, we shall consider the simulation of several example processes in both ep and pp collisions. First, we consider the diffractive structure function $F_2^{D(3)}$ as measured by the H1 Collaboration at HERA \cite{H1diff}. The complete list of modifications to the \herwig main program in order to generate pomeron exchange events using the H1 pomeron structure are as follows :

\newpage
\texttt{PROGRAM HWIGPR \\
C---BEAM PARTICLES \\
      PART1='E+      ' \\
      PART2='E-      '  ! diffracted proton \\
C---BEAM MOMENTA \\
      PBEAM1=27.6 \\
      PBEAM2=820. \\
      CALL HWIGIN \\
      MODPDF(2)=-1 \\
      iproc=19000   ! DIS hard subprocess \\
      NSTRU=6	    ! selects H1 pomeron structure routine \\
      Q2WWMN=1.E-6  ! minimum $t$ \\
      Q2WWMX=4.     ! maximum $t$ \\
      YWWMIN=.00299 ! minimum $\xpom$ \\
      YWWMAX=.00301 ! maximum $\xpom$ \\
      Q2MIN=11.9    ! minimum $Q^2$ of DIS photon \\
      Q2MAX=12.1    ! maximum $Q^2$ of DIS photon \\}

The event record is shown in table 3. The diffracted proton is represented by entry 5, labelled \texttt{E-}. Entry 4, labelled \texttt{GAMMA} is the pomeron. The incoming positron, entry 6, scatters off a \texttt{UBAR} quark in the pomeron, entry 7. The complete $F_2^{D(3)}$ as generated by \pomwig with the default parameters is compared to the H1 measurement \cite{H1diff} in figure 2. \pomwig fails only in the highest $\beta$ bins, that is at low diffractive masses, because the contribution from diffractively produced mesons is not included in \herwig. 

\begin{figure}
\centering
\begin{picture}(450,400)

\put( 0,0){\epsfig{file=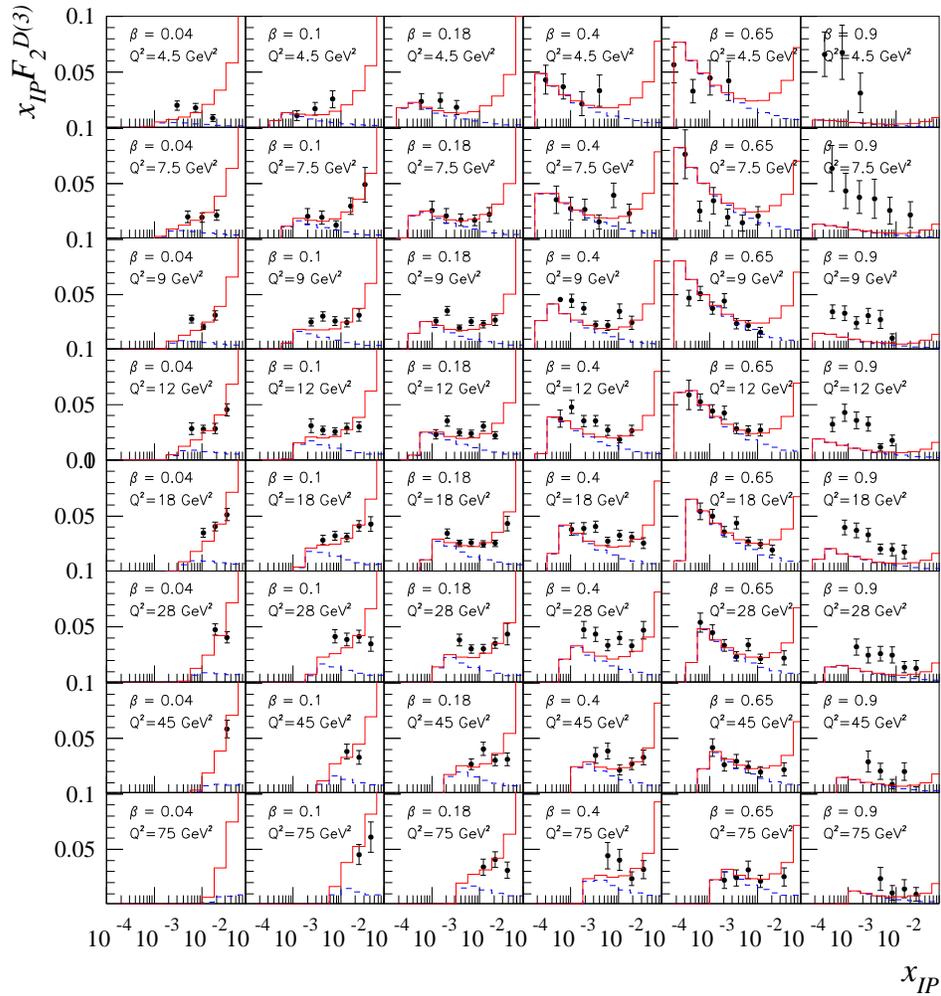,width=1.\textwidth,}}
\end{picture}
\caption{The diffractive structure function $F_2^{D(3)}$ as generated by \pomwig compared to the data of the H1 Collaboration \cite{H1diff}. The blue dashed line shows the pomeron contribution, and the red solid line shows the sum of the pomeron and reggeon contributions.}
\label{f2d3fig}
\end{figure}

In PP collisions at the Tevatron, we consider two example processes, diffractive dijet production and Higgs production via pomeron-pomeron fusion. The single diffractive process $P \bar P \rightarrow PX$ is shown in table 4. For this process, \herwig parameter \texttt{iproc} should be set to 11500. If diffractive $W$ production were required, with the $W$ decaying to electrons only, \texttt{iproc} would be set to 11451, and so on.
   The pomeron is represented in the event record by entry 4, labelled \texttt{GAMMA}. $\xpom$ for this event is 0.0067. Entry 5 in the event record, labelled \texttt{E+}, is the scattered proton. In the hard subprocess, entry 6 is the parton from the pomeron, and entry 7 is the parton from the proton. $\beta$ for this event is 0.22. 

The double diffractive process $P \bar P \rightarrow P \bar P H$, for the Higgs decaying to $b \bar b$ is shown in table 5. \texttt{iproc=11605} selects Higgs production with the Higgs decaying only to $b \bar b$. In this case, the two pomerons are entries 4 and 6 in the event record, labelled \texttt{GAMMA}. Entries 5 and 7 are the intact scattered protons, and entries 8 and 9 are the partons from the pomerons which enter the Higgs production hard subprocess.   

\begin{table}
\begin{footnotesize}
\label{disevent}
 EVENT       1:   27.60 GEV/C E+       ON   820.00 GEV/C E-       PROCESS: 19000 \\
 SEEDS:      313759      637835   STATUS:  100 ERROR:   0  WEIGHT:  1.4238E-04 \\

\begin{center}
                            ---INITIAL STATE--- \\
\vspace{.5cm}
\begin{tabular}{|l|l|l|l|l|l|l|l|l|l|l|l|l|} \hline

 IHEP  &  ID &     IDPDG& IST& MO1& MO2& DA1 &DA2 &  P-X   &  P-Y  &   P-Z \\ \hline
    1& E+  &         -11& 101 &  0 &  0 &  0 &  0 &   0.00 &   0.00 &  27.6   \\ \hline
    2& E-        &    11& 102 &  0 &  0 &  4 &  5  &  0.00 &   0.00& -820.0  \\ \hline
    3& CMF    &        0& 103  & 1 &  4&   0  & 0   & 0.20   & 0.41&   25.1  \\ \hline
    4& GAMMA    &     22  & 3   &2&   0  & 0 &  0 &   0.20  &  0.41 &  -2.5   \\ \hline
    5& E-    &        11  & 1 &  2 &  0 &  0 &  0 &  -0.20 &  -0.41& -817.5  \\ \hline

\end{tabular}
                           ---HARD SUBPROCESS---   \\

\vspace{.5cm}
\begin{tabular}{|l|l|l|l|l|l|l|l|l|l|l|l|l|} \hline

 IHEP &   ID   &   IDPDG& IST& MO1& MO2& DA1& DA2 &  P-X  &   P-Y  &   P-Z  \\ \hline
    6& E+         &  -11& 121  & 8 &  6&  11&   6   & 0.00  &  0.00  & 27.6  \\ \hline 
    7& UBAR       &   -2& 122  & 8 & 10&  12 & 10 &   0.05  &  0.10  & -0.6   \\ \hline
    8& HARD       &    0& 120  & 6 &  7&   9 & 10 &   0.07  &  0.11  & 27.0  \\ \hline
    9& E+         &  -11& 123   &8 &  9&  14 &  9 &   3.06  &  0.76  & 22.8  \\ \hline
   10& UBAR       &   -2& 124  & 8 &  7&  15 &  7 &  -3.01  & -0.66  &  4.2   \\ \hline
   11& Z0/GAMA*   &   23&   3  & 6 &  8&   0 &  0 &  -3.06  & -0.76  &  4.8  \\ \hline

\end{tabular}
\caption{Extract from a typical event record for diffractive DIS at HERA}
\end{center}
\end{footnotesize}
\end{table}

\begin{table}
\begin{footnotesize}
\label{event}
  EVENT      1:  900.00 GEV/C E+       ON  900.00 GEV/C P         PROCESS: 11500 \\
  SEEDS:     313759   637835   STATUS: 100  ERROR:   0  WEIGHT: 0.1482E+07 \\
\begin{center}
                            ---INITIAL STATE--- \\
\vspace{.5cm}
\begin{tabular}{|l|l|l|l|l|l|l|l|l|l|l|l|l|} \hline
  IHEP  &  ID &   IDPDG& IST& MO1& MO2& DA1& DA2&    P-X&     P-Y&     P-Z\\ \hline 
    1 & E+ &         -11& 101&   0  & 0 &  4 &  5  &  0.00 &   0.00&  900.00 \\
    2 & P      &    2212& 102&   0 &  0  & 0&   0 &   0.00 &   0.00& -900.00 \\
    3 &  CMF   &        0& 103 &  4   &2&   0  & 0&   -0.18 &   0.22& -893.98\\
    4 &GAMMA   &     22&   3 &  1 &  0&   0 &  0&   -0.18 &   0.22&    6.02\\ \hline
    5 &E+     &     -11&   1&   1 &  0 &  0 &  0  &  0.18&   -0.22 & 893.98\\ \hline 
\end{tabular}
                           ---HARD SUBPROCESS---   \\

\vspace{.5cm}
\begin{tabular}{|l|l|l|l|l|l|l|l|l|l|l|l|l|} \hline

  IHEP  &  ID &   IDPDG& IST& MO1& MO2& DA1& DA2&    P-X &    P-Y  &   P-Z \\ \hline  
    6 &GLUON &       21& 121&   8 &  7  &11 &  9  & -0.04 &   0.05 &   1.35 \\
    7 &GLUON &       21& 122 &  8 & 10 & 15&   6 &   0.00&    0.00 & -55.61 \\
    8 &HARD &         0& 120&   6&   7 &  9 & 10 &   0.09 &   0.03&  -54.25 \\
    9 &GLUON &       21& 123 &  8  & 6  &20&  10 &   4.12&   -0.99&   -2.04 \\
   10& GLUON     &   21& 124 &  8&   9  &26&   7  & -4.16  &  1.04&  -52.22 \\ \hline
\end{tabular}
\caption{Extract from a typical event record for the single diffractive process at Tevatron Run I}
\end{center}
\end{footnotesize}
\end{table}
\begin{table}
\begin{footnotesize}
\label{event2}
  EVENT      1: 1000.00 GEV/C E+       ON 1000.00 GEV/C E-        PROCESS: 11605 \\
  SEEDS:    9876759    63576835   STATUS: 100  ERROR:   0  WEIGHT: 0.5181E-08 \\
\begin{center}
                            ---INITIAL STATE--- \\
\vspace{.5cm}
\begin{tabular}{|l|l|l|l|l|l|l|l|l|l|l|l|l|} \hline
 
  IHEP   &  ID &   IDPDG& IST& MO1& MO2& DA1& DA2 &   P-X  &   P-Y  &   P-Z \\ \hline
    1& E+ &         -11& 101&   0 &  0  & 4&   5 &   0.00 &   0.00& 1000.00 \\ 
    2& E-  &         11& 102&   0 &  0 &  6&   7  &  0.00 &   0.00&-1000.00 \\
    3& CMF &          0& 103 &  4 &  6  & 0&   0 &   0.42 &  -0.24 &  56.61 \\
    4& GAMMA   &     22 &  3  & 1 &  0  & 0 &  0&    0.04 &   0.17 &  98.46 \\ 
    5& E+   &       -11&   1  & 1&   0 &  0  & 0  & -0.04&   -0.17 & 901.54 \\ 
    6& GAMMA  &      22 &  3  & 2   &0&   0 &  0 &   0.38 &  -0.41 & -41.85 \\
    7& E-    &       11&   1  & 2   &0&   0  & 0&   -0.38 &   0.41& -958.15 \\ \hline
\end{tabular}
                           ---HARD SUBPROCESS---   \\
\vspace{.5cm}
\begin{tabular}{|l|l|l|l|l|l|l|l|l|l|l|l|l|} \hline

  IHEP  &  ID  &  IDPDG& IST& MO1& MO2& DA1 &DA2  &  P-X &    P-Y  &   P-Z \\ \hline 
    8& UBAR   &      -2& 121 & 10 &  9 & 11  & 9  &  0.03&    0.15 &  82.80 \\
    9& UQRK   &       2& 122&  10 &  8 & 13 &  8  &  0.33 &  -0.36 & -36.53 \\  
   10& HIGGS  &      25& 120  & 8 &  9  &25&  25 &   2.35 &   0.58  & 46.29 \\  \hline
\end{tabular}
                        ---STRONG HADRON DECAYS--- \\
\vspace{.5cm}
\begin{tabular}{|l|l|l|l|l|l|l|l|l|l|l|l|l|} \hline
  IHEP  &  ID &   IDPDG& IST& MO1& MO2& DA1 &DA2 &   P-X  &   P-Y &    P-Z \\ \hline
   25& HIGGS &       25& 195 & 10 & 25  &31 & 32&    2.35 &   0.58 &  46.29 \\ \hline
\end{tabular}
                         ---H/W/Z BOSON DECAYS---  \\
\vspace{.5cm}
\begin{tabular}{|l|l|l|l|l|l|l|l|l|l|l|l|l|} \hline

  IHEP  &  ID   & IDPDG& IST& MO1& MO2& DA1& DA2  &  P-X &    P-Y &    P-Z \\ \hline 
   31& BQRK  &        5& 123&  25&  32  &33  &32 & -47.94 &  -1.24&   48.97 \\
   32& BBAR    &     -5& 124 & 25 & 31 & 35  &31  & 50.29  &  1.82  & -2.69 \\ \hline  
\end{tabular}
\caption{Extract from a typical event record for Higgs production via pomeron-pomeron fusion at Tevatron Run II}
\end{center}
\end{footnotesize}
\end{table}

\end{document}